\documentclass[%
 reprint,
aip,jcp
]{revtex4-1}

\usepackage{graphicx}
\usepackage{dcolumn}
\usepackage{bm}

\usepackage{epsfig} 
\usepackage{lipsum}

\usepackage{amsmath}
\usepackage{amsfonts}
\usepackage{amssymb}
\usepackage{mathdots}
\usepackage{mathdots}
\usepackage[utf8]{inputenc}
\usepackage{enumerate}
\usepackage{amsmath}
\usepackage{amssymb}
\usepackage{centernot}
\usepackage[mathscr]{euscript}
\usepackage{url}
\usepackage{hyphenat}
\usepackage{bm}
\usepackage{verbatim}
\usepackage{accents}
\usepackage{graphicx}
\usepackage{subcaption}
\usepackage{listings}
\usepackage{xcolor}

\usepackage{graphicx}
\usepackage{grffile}
\usepackage{algorithm,algorithmicx}
\usepackage{algpseudocode}
\algnewcommand\algorithmicinput{\textbf{Input: }}
\algnewcommand\algorithmicoutput{\textbf{Output: }}
\usepackage{afterpage}
\usepackage{epstopdf}

\everymath{\displaystyle}
\DeclareMathOperator*{\argmin}{argmin}
\DeclareMathOperator*{\argmax}{argmax}

\usepackage{lineno,hyperref}

\modulolinenumbers[5]

\newcommand{\review}[1]{\textcolor{black}{#1}}

\begin{document}


\title{Multi-fidelity machine-learning with uncertainty quantification
and Bayesian optimization for materials design: Application to ternary
random alloys}

\author{Anh Tran} 
 \thanks{\label{CorrAuth}Corresponding authors:  \href{anhtran@sandia.gov}{anhtran}, \href{jtranch@sandia.gov}{jtranch}@sandia.gov. \\
 A.T. and J.T. contributed equally to this work.}
 \affiliation{Optimization and Uncertainty Quantification, Center for Computing Research, Sandia National Laboratories}

\author{Julien Tranchida} 

\affiliation{%
 Computational Multiscale, Center for Computing Research, Sandia National Laboratories
}%

\author{Tim Wildey}
 \affiliation{Optimization and Uncertainty Quantification, Center for Computing Research, Sandia National Laboratories}

\author{Aidan P. Thompson}%
\affiliation{%
 Computational Multiscale, Center for Computing Research, Sandia National Laboratories
}%

\date{\today}

\begin{abstract}
  We present a scale-bridging approach based on a 
  multi-fidelity (MF) machine-learning (ML) framework leveraging 
  Gaussian processes (GP) to fuse atomistic computational model
  predictions across multiple levels of fidelity.
  Through the posterior variance of the MFGP, our framework 
  naturally enables uncertainty quantification, providing estimates 
  of confidence in the predictions. 
  We used Density Functional Theory as high-fidelity prediction, while a 
  ML interatomic potential is used as low-fidelity prediction.
  Practical materials design efficiency is demonstrated by reproducing the
  ternary composition dependence of a quantity of interest
  (bulk modulus) across the full aluminum-niobium-titanium ternary random 
  alloy composition space.
  The MFGP is then coupled to a Bayesian optimization procedure
  and the computational efficiency of this approach is demonstrated
  by performing an on-the-fly search for the global optimum of bulk
  modulus in the ternary composition space. 
  The framework presented in this manuscript is the first application of MFGP 
  to atomistic materials simulations fusing predictions between Density 
  Functional Theory and classical interatomic potential calculations.
\end{abstract}

\pacs{Valid PACS appear here}
\maketitle


\section{Introduction}
\label{sec:Introduction}


Materials design can be seen as an inverse problem
in the structure-property relationship. 
In the context of modern random alloys, such as medium- and 
high-entropy alloys, optimizing functional performance
can require the exploration of vast 
composition spaces.
It is therefore highly desirable to develop numerical tools to
enable the prediction of optimal concentrations of the different 
components, in order to optimize a set of materials  
properties~\cite{gubaev2019accelerating,kostiuchenko2019impact}.
Such predictive tools could greatly reduce experimental testing and 
manufacturing costs, and are of great interest~\cite{wen2019machine}. 


High-accuracy \emph{first-principles} approaches, such as Density
Functional Theory (DFT), are an efficient tool to complement
experiments in this optimization process.
However, their predictive capabilities are limited by their 
computational cost and ${\cal O}(N^3)$ scaling
\cite{guerra1998towards}, making them impractical 
for the optimization of multiple properties across large composition 
spaces, particularly for properties requiring more than a few hundred 
atoms to resolve.


Machine learning (ML) interatomic potentials (IAPs) are emerging as a 
promising solution to run large-scale problems, while preserving a 
level of accuracy close to \emph{first-principles}
methods~\cite{bartok2010gaussian}.
Trained on extensive datasets of atomic configurations (usually
extracted from DFT calculations), ML-IAPs are computationally efficient 
and preserve the ${\cal O}(N)$ linear
scaling of computational cost with atom count \cite{zuo2020performance}.
This leads to prediction costs orders of magnitude cheaper than \emph{ab
initio} calculations, allowing to effectively bridge the gap between 
nano- and meso-scale models with a controlled level of 
accuracy~\cite{chen2017accurate,wood2019data}.


However, MD performed with ML-IAPs loses some of its prediction 
accuracy (and thus  materials design and discovery capabilities) 
when configurations and compositions departing from the training set 
are considered.
While a lot of effort has been concentrated on training and 
validating ML-IAPs, 
the use of DFT and MD predictions mostly remains segregated:
\review{in the context of ML-IAPs, the former is used to build training 
and testing sets, and the latter is used to run statistical averaging 
and large scale simulations.}
Although they form a natural pair of high- and low-fidelity  
computational models, there seems to be a lack of research effort to 
fuse their information.


In the field of materials science, few studies have sought to
exploit the correlation between low- and
high-fidelity levels to obtain high-accuracy multi-fidelity (MF) 
predictions, thus fusing the information across multiple 
levels of fidelity hierarchically.
Batra \emph{el al.} recently wrote a comprehensive review
of different levels of fidelity used to obtain forces in atomistic 
materials modeling\cite{batra2020machine}.
Pilania \emph{el al.} developed a MF co-kriging ML framework 
enabling low-cost accurate quantum mechanical predictions of bandgaps 
by fusing the information obtained from DFT calculations performed
with different exchange correlation functionals \cite{pilania2017multi}. 
Other recent studies have been aiming at fusing information and
predictions across multiple-levels of fidelity for materials
study and design 
\cite{acar2018crystal,sun2010two,razi2018fast,chen2020multi}.

Gaussian processes (GP) and associated Bayesian optimization (BO) 
methods are among popular approaches that have been employed extensively
in computational materials science~\cite{ramprasad2017machine},
most notably in the construction of Gaussian Approximation Potential 
(GAP) ML-IAPs \cite{bartok2010gaussian}.
This includes studies of shape memory alloy \cite{xue2016accelerated},
polymer dielectrics \cite{mannodi2016machine}, polymer
bandgaps \cite{patra2020multi}, dopant formation energies in hafnia
\cite{batra2020machine}, search for saddle points on high-dimensional 
potential energy surfaces
\cite{tran2018gpdft,tran2018efficient,koistinen2019minimum},
fusing MD simulation with different time-steps \cite{razi2018fast}, and 
calibrating coarse-grain MD potential \cite{razi2020force}.

Our previous work developed a data-driven MF-ML method leveraging a
multi-fidelity Gaussian-process (MFGP) approach 
\cite{tran2019sbfbo2cogp,tran2020smfbo2cogp}
and utilized it to fuse the predictions of the same quantity of interest 
(QoI) from multiple levels of fidelity. 
Here we apply the approach to atomistic materials simulation.
The high- (HF) and low-fidelity (LF) models are 
DFT and \review{a Spectral Neighbor Analysis Potential (SNAP)} 
ML-IAP~\cite{thompson2015spectral}, respectively. 
A MF Bayesian optimization is then applied to optimize the QoI with 
respect to chemical composition. 
The goal of this work is to demonstrate the applicability of our 
MF framework to applications in atomistic materials simulation.
To the knowledge of the authors, this work presents the first 
MF-ML framework exploiting the respective strengths of DFT and ML-IAP
(i.e. high-accuracy but expensive predictions versus low-cost and 
scalable but less accurate predictions) and to fuse their predictions.

This manuscript is organized as follows. 
In Section \ref{sec:Simulations}, we describe the details of DFT 
simulations (Section \ref{subsec:DFTsimulations}) and MD simulations 
(Section \ref{subsec:MDsimulations}) for AlNbTi system. 
Section \ref{sec:MFGP} briefly formulates the MFGP framework and its MF 
BO extension. 
Section \ref{sec:Results} presents the results for AlNbTi system. 
Section~\ref{sec:Conclusion} discusses and concludes the paper.


\section{High- and Low-fidelity bulk modulus calculations}
\label{sec:Simulations}

In this section, we describe the \emph{ab initio} and classical 
calculations used to build the high- and low-fidelity models, 
respectively. 
The high-fidelity model relies on Density Functional Theory (DFT)
and is presented in section~\ref{subsec:DFTsimulations}.
The low-fidelity model is based on a ML-IAP and is presented 
in~\ref{subsec:MDsimulations}.

As a proof of concept calculation, the bulk modulus is chosen to
be the physical quantity of interest for the multi-fidelity prediction.
It is known to converge more rapidly with respect to the
k-point sampling than other elastic properties, such as the shear 
moduli~\cite{mattsson2004designing},
which made our high-fidelity calculations more tractable. 
However, our MF approach is easily transferable to any other
QoI that can be evaluated by both \emph{ab initio} and classical
calculations.

\subsection{Density functional theory}
\label{subsec:DFTsimulations}

\emph{Ab initio} calculations were carried out using plane-wave 
density functional theory as implemented in the Quantum ESPRESSO 
package \cite{giannozzi2009quantum,giannozzi2017advanced}, within the 
framework of the PBE formalism \cite{perdew1996generalized}. 
The interactions between the electrons and ions were represented using 
projector augmented wave (PAW) pseudopotentials.
We used a kinetic energy cutoff of 55 Ry for the wave function and 600 Ry 
for the charge density. 
The Brillouin zone was sampled using a 2x2x2 k-point grid and
Gaussian smearing with a smearing value of 0.025 eV.
For all calculations, the convergence threshold for
self-consistency was set to 10$^{-8}$.
\review{
Within this \emph{ab initio} formalism and for a large amount of chemical
compositions and materials, bulk mudulus calculations have been shown to 
remain within a 15\% accuracy compared to experimental 
measurements \cite{tran2016rungs}.  
This level of accuracy proved sufficient for materials design and
discovery \cite{saal2013materials}. 
}

Initial body-centered cubic (bcc) cells are generated following the
approach \review{described in former work~\cite{tranchida2020AlNbTi_SNAP}.}
The atoms of the cell are randomly selected to be Al, Nb, or Ti, subject 
to the constraint of a particular total number of atoms of each element.
The number of distinct chemical compositions that can be generated 
with configurations of $N$ atoms and $E$ elements is given by the 
multiset coefficient $\left (( \tiny\begin{array}{c} N \\ E \end{array} ) \right ) = \left ( \tiny\begin{array}{c} N+E-1 \\ E-1 \end{array} \right )$.
The total number of distinct compositions achievable with 54 atoms and 
3 elements is then $1540$.
The number of expensive DFT calculations to be performed is reduced by
considering atom counts that are multiples of 3, leading to 190 distinct
compositions.
For each composition, one random coloring is generated. 
It is shown in Ref.~\cite{tranchida2020AlNbTi_SNAP} that for 54 atom 
cells, the particular random coloring does not strongly modify the bulk modulus
calculation. We leverage that result to reduce the number of expensive 
calculations to be performed.
For each of these 190 structures, an equation of state (EOS) is computed 
by ranging its 
volume through eight points. The volume variation corresponds to an
approximate compression and expansion of $6\%$.
Fig.~\ref{fig:cropped_configs} displays the results of eight of those
EOS calculations at the equicomposition point.

The EOS results were interpolated with a Birch-Murnaghan equation, 
using a third order
polynomial~\cite{birch1978finite,murnaghan1937finite}. 
This procedure allowed us to extract the high-fidelity bulk modulus
corresponding to a given composition.

\subsection{Classical potential calculations}
\label{subsec:MDsimulations}

Classical potential calculations were carried out using the LAMMPS 
package~\cite{plimpton1995fast} and a SNAP machine learning
potential~\cite{thompson2015spectral} whose training set contained
information at the equicomposition point, and for each of the
single-element phases.
Details of the training and testing of the SNAP potential are provided in
Tranchida \emph{et al.} \cite{tranchida2020AlNbTi_SNAP}. 

The bulk modulus is computed following the same approach as
described in section~\ref{subsec:DFTsimulations} above.
\review{
The accuracy of bulk modulus calculations performed with a classical ML-IAP
should be expected to be at best as good as the HF DFT model (and
commonly lower), as the potential is trained to match this reference 
information.
}
As the computational cost of the LF 
calculations is very small compared to the HF calculations, we densified the
number of LF points in three ways.
First, we computed the LF bulk modulus for all 1540 possible
compositions of 54 atom cells.
Then, instead of varying the cell volume through eight points,
we use 100 points. 
This ensures a better convergence of the Birch-Murnaghan 
polynomials.
Finally, for each composition point, the bulk modulus is averaged
over 10 initial cells, each one corresponding to different random 
coloring for the given composition.

In this proof-of-concept study, we decide to keep the same cell size
across the two levels of fidelity (54 atoms).
This simplified the comparison between the LF and HF models. 
However, it is straightforward and almost computationally
transparent to largely increase the LF cells. 
This could for example enable the discovery of correlations
between large scale effects and information extracted from small
DFT cells.
This path would increase the scale-bridging aspect of our work, and will be
explored in future studies.

\begin{figure}[!htbp]
\centering
\includegraphics[width=0.99\columnwidth,keepaspectratio]{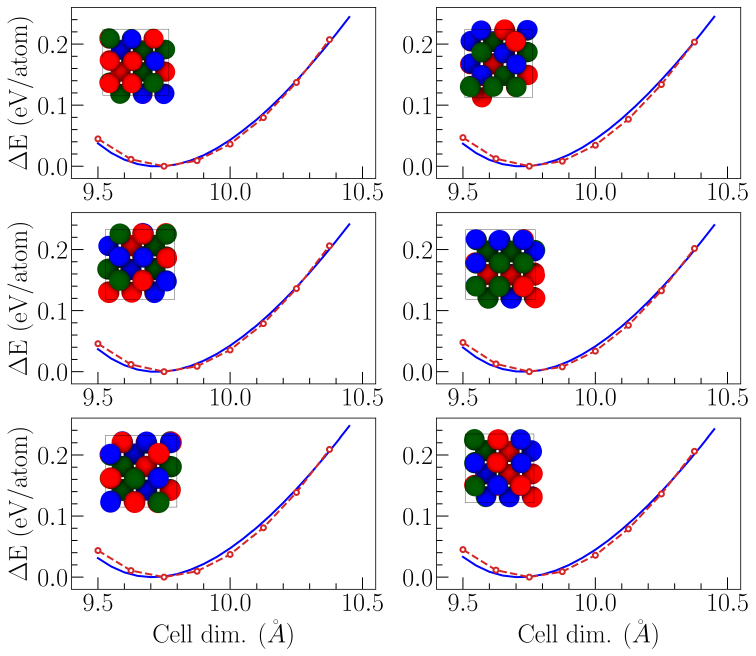}
\caption{Equation of state calculations for six configurations.
  The red circles and blue lines display the high-fidelity DFT
  and low-fidelity SNAP results, respectively.}
\label{fig:cropped_configs}
\end{figure}

Fig.~\ref{fig:cropped_configs} displays the comparison of eight LF
EOS calculations compared to the HF ones, at the equicomposition point.
As detailed in Tranchida \emph{et al.}~\cite{tranchida2020AlNbTi_SNAP}, 
about $70\%$ of the training set of the SNAP potential contains 
equicomposition information. 
This explains the excellent agreement displayed on
Fig.~\ref{fig:cropped_configs}.
Inferior agreement is expected when large departures from the
equicomposition point are observed.


\section{Multi-fidelity Gaussian-process and multi-fidelity Bayesian optimization}
\label{sec:MFGP}

In this section, we briefly review the formulation of GP, particularly the MF extension of the classical GP, and the relevant MF BO method \cite{tran2020smfbo2cogp,tran2019sbfbo2cogp}.

\subsection{Gaussian Processes}

Gaussian process regression is an efficient and flexible framework to approximate a response surface for a single-fidelity, single-objective function. 
We briefly summarize the GP theoretical formulation for the sake of completeness \cite{shahriari2016taking}. 
Let $\mathcal{D} = (\bm{x}_i, y_i)_{i=1:n}$ denote the dataset of $n$ observations with output $y$ and $d$-dimensional input $\bm{x} \in \mathcal{X} \subseteq \mathbb{R}^d$. 
A GP is a nonparametric model characterized by its prior mean function $\mu_0(\bm{x}): \mathcal{X} \to \mathbb{R}$ and a covariance function $k(\bm{x},\bm{x'}): \mathcal{X} \times \mathcal{X} \to \mathbb{R}$. 
Assuming that the observations $\bm{f}=f_{1:n}$ are jointly Gaussian, and the observation $y$ is normally distributed given $\bm{f}$, i.e. 
\begin{equation}
\bm{f} | \bm{x} \sim \mathcal{N}(\bm{m}, \bm{K}),
\end{equation} 
\begin{equation}
\bm{y} | \bm{f},\sigma^2 \sim \mathcal{N}(\bm{f}, \sigma^2 \bm{I}),
\end{equation} 
where $m_i:=\mu(\bm{x}_i)$ and $K_{i,j}:= k(\bm{x}_i, \bm{x}_j)$. 

The classical GP regression formulation assumes a stationary covariance matrix and only considers the weighted distance $r^2(\bm{x}, \bm{x'}) = (\bm{x} - \bm{x'})^T \Lambda (\bm{x} - \bm{x'})$, where $\Lambda$ is a diagonal matrix of $d$ squared length scales $\theta_i^2$. 
Mat{\'e}rn kernels offer a broad class for stationary kernels, controlled by a smoothness parameter $\nu>0$ (cf. Section 4.2, \cite{rasmussen2006gaussian}), including the square-exponential ($\nu \to \infty$) and exponential $(\nu = 1/2)$ kernels widely used in the literature. 
The $\nu = 3/2$ Mat{\'e}rn kernel $k(\bm{x}, \bm{x'}) = \theta_0^2 \exp{(-\sqrt{3}r)} (1+\sqrt{3} r)$ is used in this work. 
At a known sampling point $\bm{x} \in \mathcal{X}$, the posterior mean $\mu (\bm{x})$ is
calculated by
\begin{equation}
\label{eq:mean}
\mu(\bm{x}) = \mu_0(\bm{x}) + \bm{k}(\bm{x})^T (\bm{K} + \sigma^2 \bm{I})^{-1} (\bm{y} - \bm{m}),
\end{equation}
and the posterior variance $\sigma^2 (\bm{x})$ is given by
\begin{equation}
  \begin{array}{lll}
    \sigma^2(\bm{x}) &=& k(\bm{x}, \bm{x}) \\
    &&- \bm{k}(\bm{x})^T (\bm{K}  + \sigma^2 \bm{I})^{-1} \bm{k}(\bm{x}),
  \end{array}
\label{eq:variance}
\end{equation}
where $\bm{k}(\bm{x})$ is a vector of covariance $\bm{k}(\bm{x})_i = k(\bm{x}, \bm{x}_i)$, $\sigma^2 = \frac{1}{n} (\bm{y} - \mu_0(\bm{x}))^T \bm{K}^{-1} (\bm{y} - \mu_0(\bm{x})) $ is the intrinsic variance. 
To obtain the hyper-parameter $\theta = (\theta_i)_{i=1:d}$, we maximize the log marginal likelihood, which is computed as
\begin{equation}
\begin{array}{lll}
\log p(\bm{y} | \bm{x}_{1:n}, \theta) &=& - \frac{1}{2} (\bm{y} - \bm{m})^T (\bm{K}^{\theta} + \sigma^2 \bm{I})^{-1} (\bm{y} - \bm{m}) \\
&& - \frac{1}{2} \log{| \bm{K}^{\theta} + \sigma^2 \bm{I} |} - \frac{n}{2} \log{(2\pi)}.
\end{array}
\end{equation}
Here, $\bm{K}^{\theta}$ is emphasized to be strongly dependent on $\theta$. 

\subsection{Multi-fidelity Gaussian Processes}

We assume that the prediction at highest level of fidelity, i.e. level 
$s$, can be written as an auto-regressive model \cite{xiao2018extended}, 
\begin{equation}
\label{eq:mfgpFormulation}
f_H(\bm{x}) = \rho f_L(\bm{x}) + \delta (\bm{x}), 
\end{equation}
where $f_H(\bm{x})$ and $f_L(\bm{x})$ denote the high- and low-fidelity predictions, respectively, $\rho$ is the scaling factor, and $\delta(\bm{x})$ is the discrepancy between the high- and low-fidelity model. 
The dataset $\mathcal{D}$ is divided into $\mathcal{D}_L = (\bm{x}_i, y_i)_{i=1:n_L}$ and $\mathcal{D}_H = (\bm{x}_i, y_i)_{i=1:n_H}$, corresponding to low-fidelity and high-fidelity datasets, respectively. 
Our multi-fidelity formulation is closely related to Couckuyt \emph{et al.}~ \cite{couckuyt2012blind,couckuyt2013oodace,couckuyt2014oodace} and Forrester \emph{et al.}~ \cite{forrester2007multi}. 
Following the auto-regressive scheme described in Equation \ref{eq:mfgpFormulation}, the main idea of MFGP in two levels of fidelity is to model $f_L$ as the first GP and the discrepancy $\delta(\bm{x})$ as the second GP, before fusing both of them together. 

In the MFGP, the covariance matrix $\tilde{K}$ is computed as
\begin{widetext}
\begin{equation}
\label{eq:covarianceMF}
\tilde{\bm{K}} = \begin{pmatrix}
\sigma_L^2  \bm{K}_L (\bm{x}_L, \bm{x}_L) & \rho  \sigma_L^2  \bm{K}_L(\bm{x}_L, \bm{x}_H) \\ 
\rho  \sigma_L^2  \bm{K}_L(\bm{x}_H, \bm{x}_L) & ~~\rho^2  \sigma_L^2  \bm{K}_L(\bm{x}_H, \bm{x}_H) + \sigma_d^2  \bm{K}_D (\bm{x}_H,\bm{x}_H)
\end{pmatrix}.
\end{equation}
\end{widetext}

At the high-fidelity level, the posterior mean $\mu (\bm{x})$ and the posterior variance $\sigma^2 (\bm{x})$ are computed, respectively, as
\begin{equation}
\begin{array}{lll}
\mu (\bm{x}) &=& \mu_0 (\bm{x}) + \tilde{\bm{k}} (\bm{x})^T (\tilde{\bm{K}} + \sigma^2 \bm{I})^{-1} (\tilde{\bm{y}} - \tilde{\bm{m}}), \\
\sigma^2 (\bm{x}) &=& \rho^2 \sigma_L^2 (\bm{x}) + \sigma_d^2(\bm{x}) \\
&& - \tilde{\bm{k}}(\bm{x})(\tilde{\bm{K}} + \sigma^2 \bm{I})^{-1} \tilde{\bm{k}}(\bm{x}), \\
\end{array}
\end{equation}
where 
\begin{equation}
\begin{array}{lll}
\mu_H (\bm{x}) &=& \rho \mu_L (\bm{x}) + \mu_d(\bm{x}),\\ \sigma_L^2(\bm{x}) &=& k_L(\bm{x},\bm{x}),\\
\sigma_d (\bm{x}) &=& k_d(\bm{x}, \bm{x}),\\
\tilde{\bm{y}} &=& 
\begin{pmatrix}
\bm{y}_L^T \\ \bm{y}_H^T 
\end{pmatrix} = 
\begin{pmatrix}
\left(y_{1,L}, \dots, y_{n_L,_L}\right)^T \\ 
\left(y_{1,H}, \dots, y_{n_H,_H}\right)^T 
\end{pmatrix}, \\
\tilde{\bm{m}} &=& 
\begin{pmatrix} \bm{\mu}_L \\ \bm{\mu}_H \end{pmatrix} \\
&=& \begin{pmatrix} 
\left( \mu_L(\bm{x}_{1,L}), \dots, \mu_L(\bm{x}^{n_L,L}) \right)^T \\
\left( \mu_H(\bm{x}_{1,H}), \dots, \mu_H(\bm{x}^{n_H,H}) \right)^T
\end{pmatrix}, \\
\tilde{\bm{k}} (\bm{x}) &=& 
\begin{pmatrix} \rho \bm{k}_L (\bm{x}) \\ \bm{k}_H (\bm{x}) \end{pmatrix} \\
&=& \begin{pmatrix} 
\left( \rho k_L(\bm{x}, \bm{x}_{1,L}), \dots, \rho k_L(\bm{x}, \bm{x}_{n_L,L}) \right)^T \\
\left(      k_H(\bm{x}, \bm{x}_{1,H}), \dots,      k_H(\bm{x}, \bm{x}_{n_H,H}) \right)^T
\end{pmatrix}\\
k_H (\bm{x}, \bm{x'}) &=& \rho^2 k_L(\bm{x}, \bm{x'}) + k_d(\bm{x}, \bm{x'}).\\
\end{array}
\end{equation}
$\bm{x}_{i,L}$ and $\bm{x}_{i,H}$ denote the $i^{\text{th}}$ inputs at the low- and high-fidelity levels, respectively. 
$y_{i,L}$ and $y_{i,H}$ denote the $i^{\text{th}}$ observations at the low- and high-fidelity levels, respectively. 
The hyper-parameters in $\tilde{\theta} = (\theta_L, \theta_H)$ in $k_L(\cdot,\cdot)$ and $k_D(\cdot,\cdot)$ can be obtained by maximizing the log marginal likelihood as
\begin{equation}
\begin{array}{lll}
&& \log p(\tilde{\bm{y}} | \bm{x}_{1:n_L}, \bm{x}_{1:n_H}, \tilde{\theta}) \\
&=& - \frac{1}{2} (\tilde{\bm{y}} - \tilde{\bm{m}})^T (\tilde{\bm{K}}^{\tilde{\theta}} + \sigma^2 \bm{I})^{-1} (\tilde{\bm{y}} - \tilde{\bm{m}}) \\
&& - \frac{1}{2} \log{| \tilde{\bm{K}}^{\tilde{\theta}} + \sigma^2 \bm{I} |} - \frac{n_H + n_L}{2} \log{(2\pi)}.
\end{array}
\end{equation}
For further details, readers are referred to previous work in literature \cite{couckuyt2013oodace,xiao2018extended,yang2020bifidelity}.

\subsection{Multi-fidelity Bayesian Optimization}
\label{subsec:AcqFunc}
In the traditional BO method, the next sampling location is determined by maximizing an acquisition function, i.e.,
\begin{equation}
\bm{x}^* = \argmax_{\bm{x} \in \mathcal{X}} \ a(\bm{x}), 
\end{equation}
where $a(\bm{x})$ denotes the acquisition function and $\bm{x}^*$ is the next sampling location. 
The acquisition function is deeply connected to an underlying utility function, which corresponds to a reward scheme for performance of the new sampling point relative to previous samples. 
There are three acquisition functions that are widely used: the probability of improvement (PI), the expected improvement (EI), and the upper-confident bounds (UCB), but other forms also exist. 

The UCB acquisition function \cite{srinivas2009gaussian,srinivas2012information,daniel2014active} is defined as
\begin{multline}
a_{\text{UCB}}(\bm{x};\{\bm{x}_i,y_i \}_{i=1}^n,\theta) = \\
\mu(\bm{x};\{\bm{x}_i,y_i \}_{i=1}^n,\theta) + \kappa \sigma(\bm{x};\{\bm{x}_i,y_i \}_{i=1}^n,\theta),
\end{multline}
where $\kappa$ is a hyper-parameter describing the acquisition exploitation-exploration balance. 
We adopt the $\kappa$ computation from Daniel \emph{et al.}~\cite{daniel2014active}, which is based on Srinivas \emph{et al.}~\cite{srinivas2009gaussian,srinivas2012information}, instead of fixing $\kappa$ as a constant. 

Regarding the fidelity selection criteria, we adopt the approach developed in \cite{tran2020smfbo2cogp} by choosing level $t^*$ such that
\begin{equation}
t^* = \argmin_t \left( C_t \int_\mathcal{X}\sigma^2(\bm{x}) d\bm{x} \right),
\end{equation}
where $C_t$ is the computational cost at level $t$, $t$ can be low- or high-fidelity level. 
The term $\int_\mathcal{X}\sigma^2(\bm{x}) d\bm{x}$ is sometimes referred to as the integrated mean square error, as opposed to the conventional mean square error $\sigma^2(\bm{x})$, to reduce the number of sampling points on the boundary of $\mathcal{X}$ for further efficiency improvement. 
To facilitate information at the high-fidelity level when the computational cost are comparable, we impose a hard condition that if $C_L n_L \geq C_H n_H$, meaning that some of the computational budget for building the low-fidelity dataset $\mathcal{D}_L$ could be traded for building the high-fidelity dataset $\mathcal{D}_H$, then the high-fidelity is chosen.


\section{Results}
\label{sec:Results}



\subsection{Fusing low- and high-fidelity bulk modulus predictions}


\begin{figure}[!htbp]

\centering
\begin{subfigure}[b]{0.50\textwidth}
	\centering
	\includegraphics[width=1\textwidth,keepaspectratio]{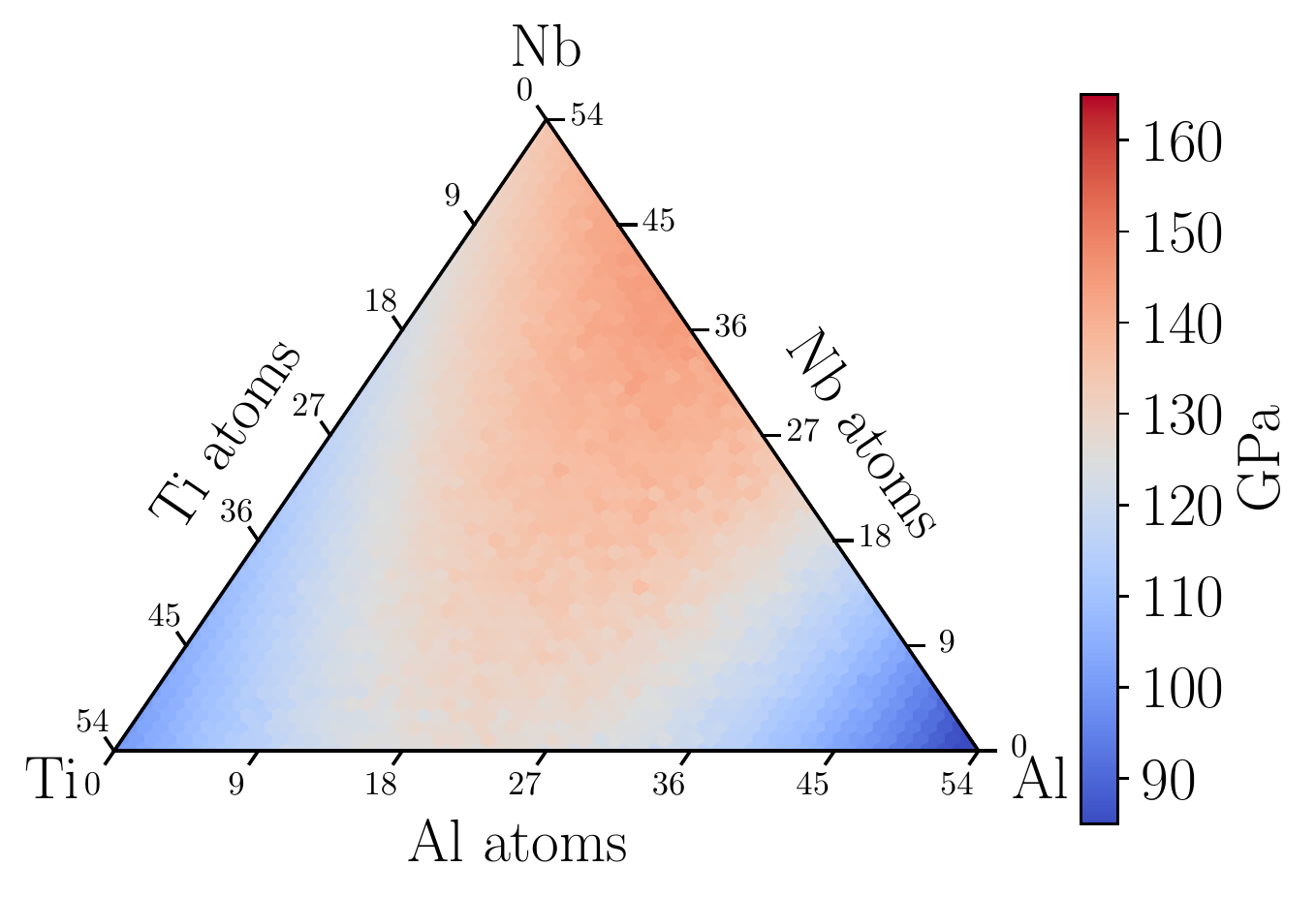}
	\caption{Low-fidelity.}
	\label{fig:cropped_lfTernary0K}
\end{subfigure}
\hfill
\begin{subfigure}[b]{0.50\textwidth}
	\includegraphics[width=1\textwidth,keepaspectratio]{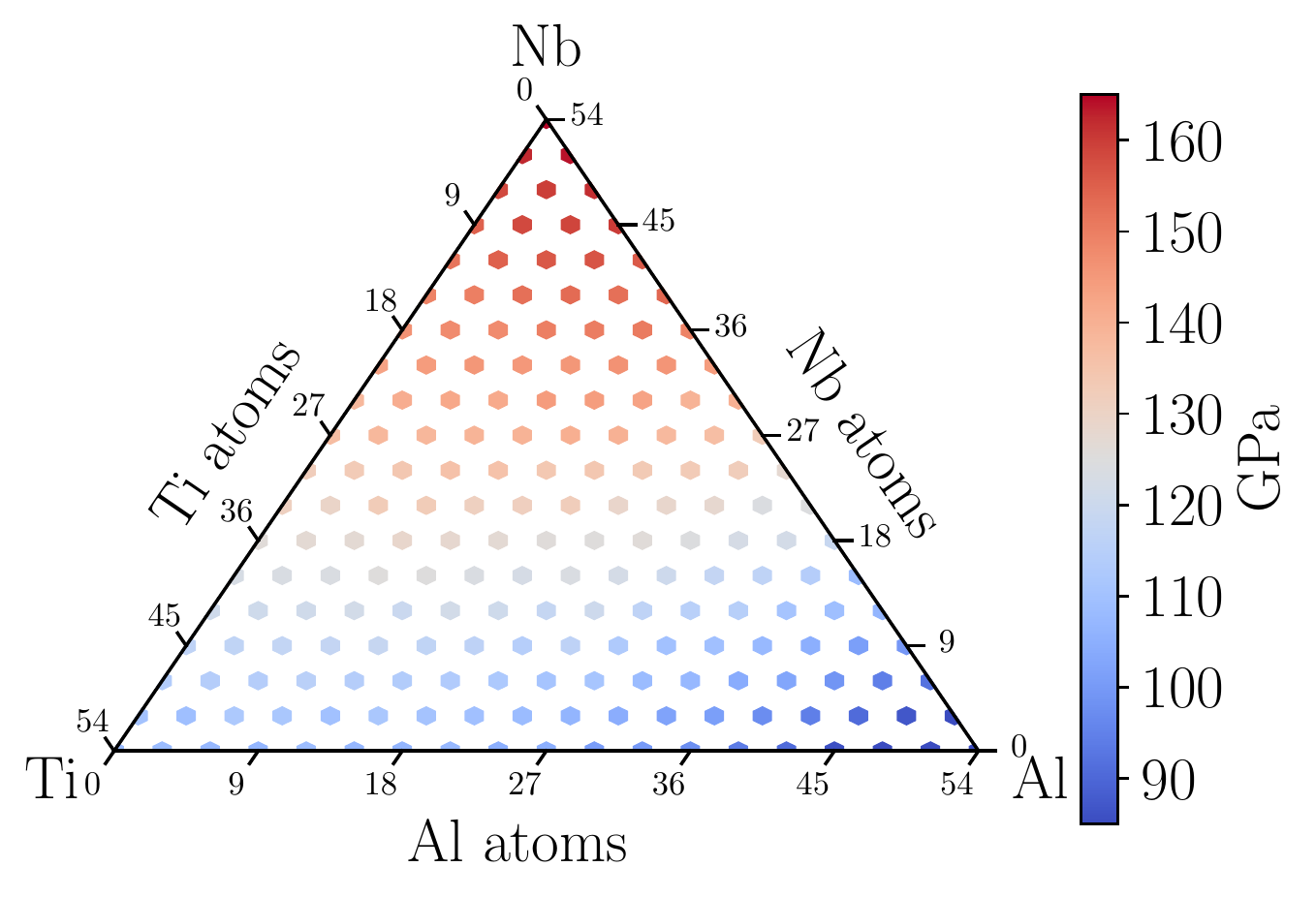}
	\caption{High-fidelity.}
	\label{fig:cropped_hfTernary0K}
\end{subfigure}
\hfill
\begin{subfigure}[b]{0.50\textwidth}
	\includegraphics[width=1\textwidth,keepaspectratio]{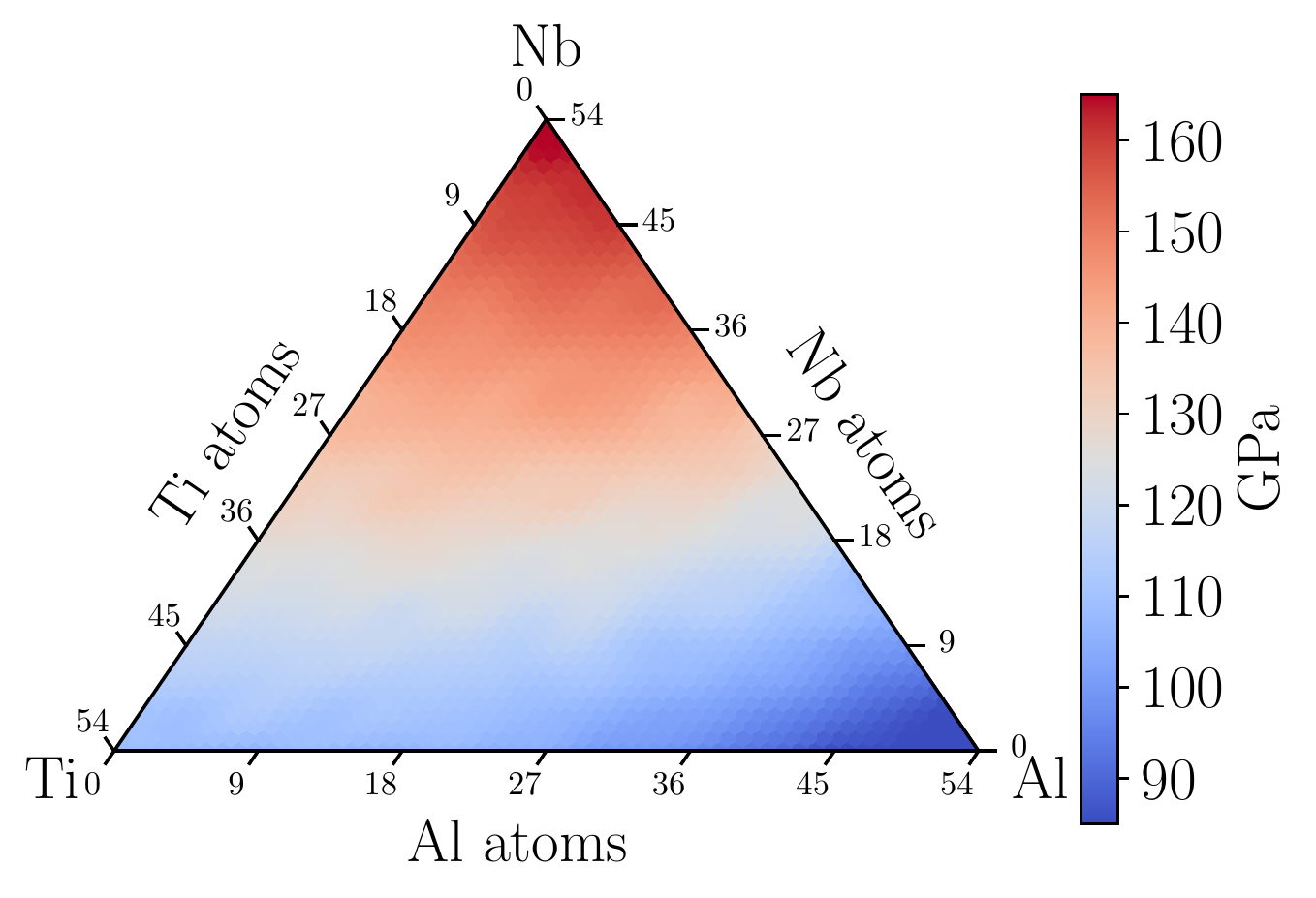}
	\caption{Multi-fidelity.}
	\label{fig:cropped_mfTernary0K}
\end{subfigure}
\vfill

\caption{Low-fidelity (SNAP potential -- Figure \ref{fig:cropped_lfTernary0K}) and 
  high-fidelity (DFT -- Figure \ref{fig:cropped_hfTernary0K}) calculated values of bulk modulus for AlNbTi ternary compositions, as well as multi-fidelity (SNAP/DFT -- Figure \ref{fig:cropped_mfTernary0K}) 
  predictions. 
}
\label{fig:LfHfMfpredictions}
\end{figure}

As explained in section~\ref{subsec:DFTsimulations} and
\ref{subsec:MDsimulations}, 190 HF and 1540 LF calculations of 
bulk modulus at different compositions were performed to build 
the dataset. Figure \ref{fig:LfHfMfpredictions} displays the LF
(Fig.~\ref{fig:cropped_lfTernary0K}) and HF 
(Fig.~\ref{fig:cropped_hfTernary0K}) data sets, as well as the MF
predictions using the MFGP approach 
(Fig.~\ref{fig:cropped_mfTernary0K}).
As can be seen by comparing Fig.~\ref{fig:cropped_hfTernary0K} and 
Fig.~\ref{fig:cropped_mfTernary0K}, 
starting from a sparse HF ternary composition map, 
the MFGP approach is able to
leverage the LF-HF correlations to produce a high-resolution diagram predicting the composition dependence of the bulk 
modulus with an accuracy almost equal to DFT.

\begin{figure*}[!htbp]

\centering
\begin{subfigure}[b]{0.30\textwidth}
	\centering
	\includegraphics[width=1\textwidth,keepaspectratio]{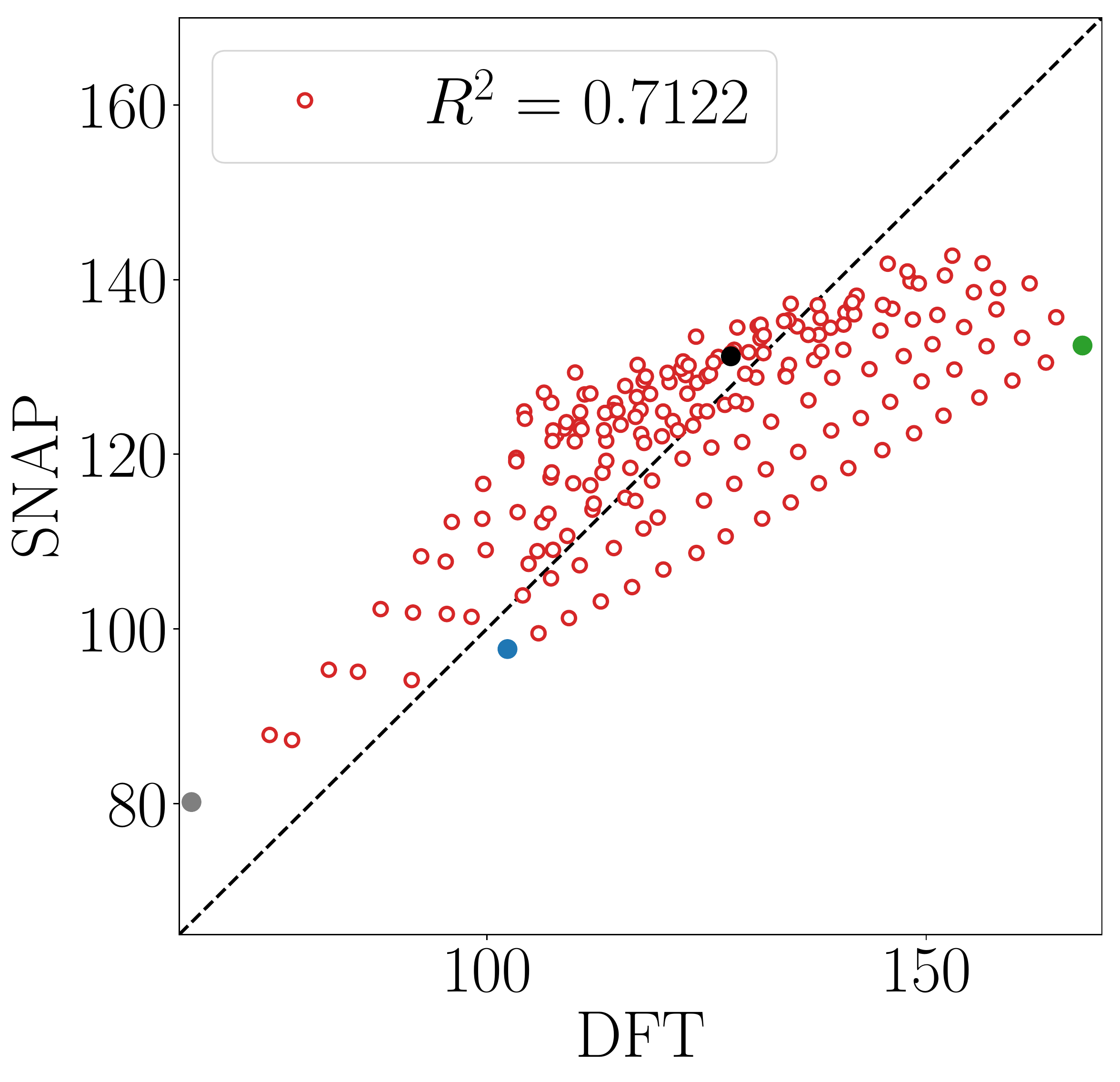}
	\caption{SNAP}
	\label{fig:cropped_predDFTvsSNAP}
\end{subfigure}
\hfill
\begin{subfigure}[b]{0.30\textwidth}
	\centering
	\includegraphics[width=1\textwidth,keepaspectratio]{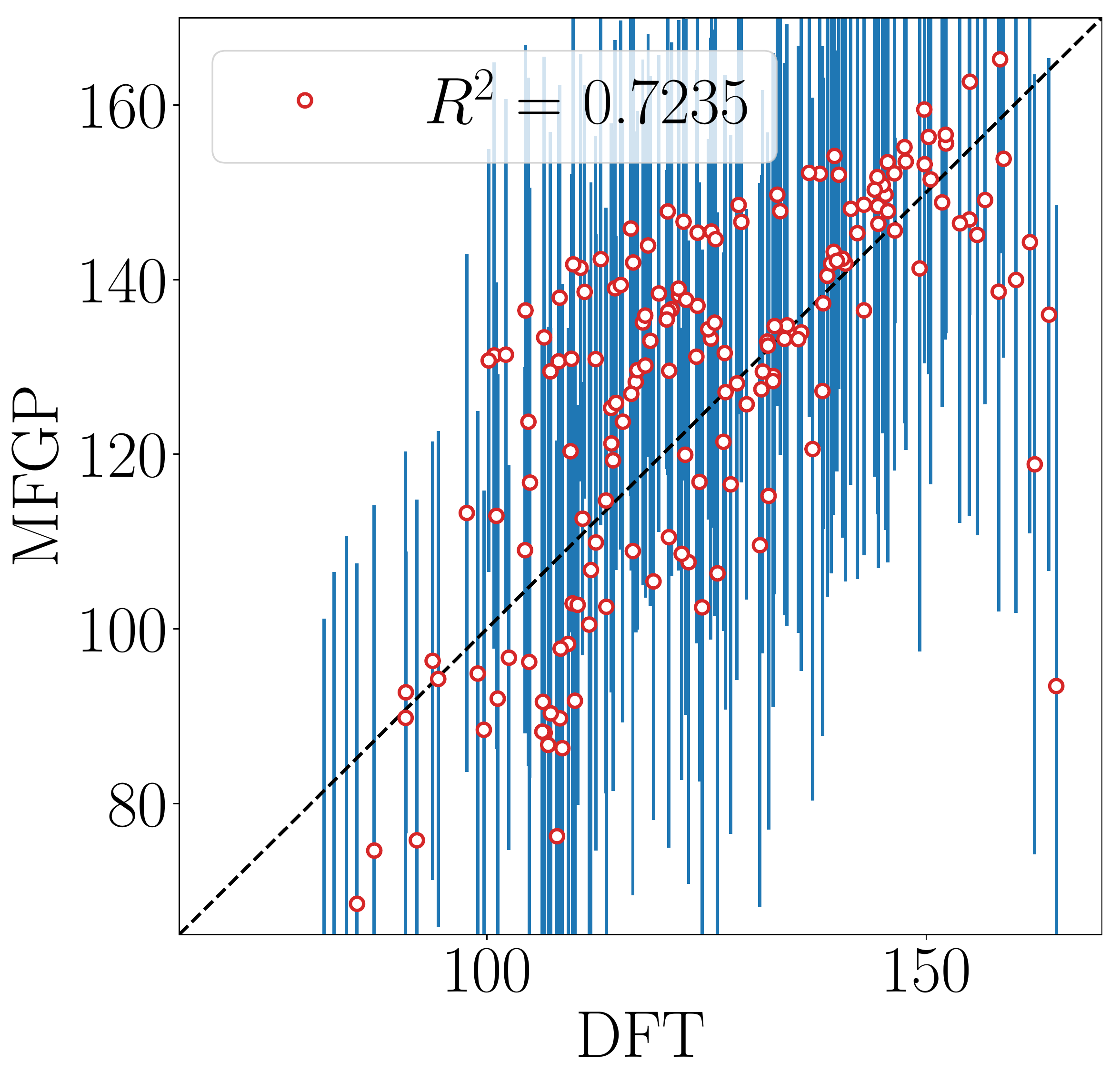}
	\caption{MFGP-10}
	\label{fig:cropped_predDFTvsMFGP-10-90}
\end{subfigure}
\hfill
\begin{subfigure}[b]{0.30\textwidth}
	\centering
	\includegraphics[width=1\textwidth,keepaspectratio]{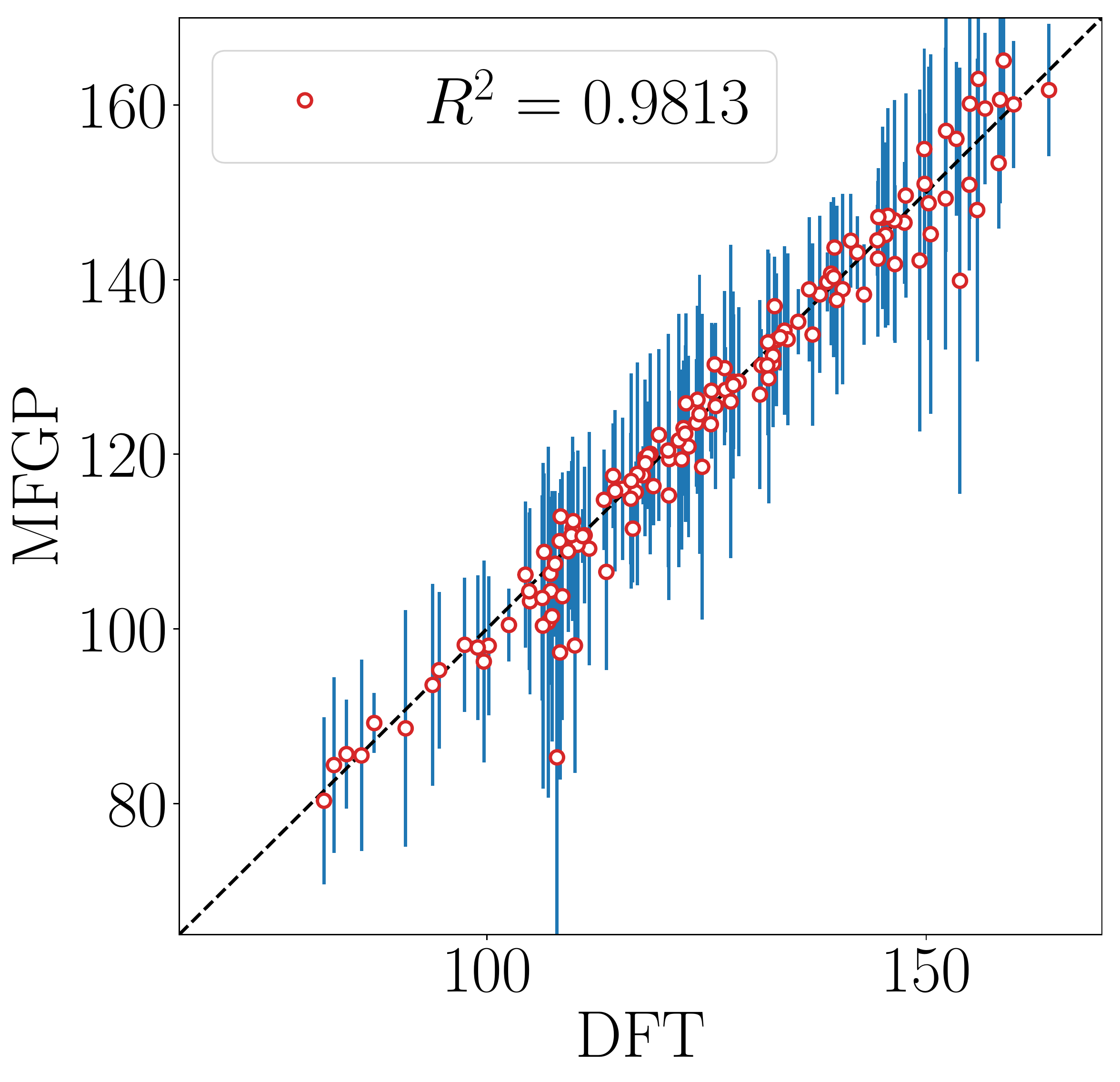}
	\caption{MFGP-30}
	\label{fig:cropped_predDFTvsMFGP-30-70}
\end{subfigure}
\vfill

\begin{subfigure}[b]{0.30\textwidth}
	\includegraphics[width=1\textwidth,keepaspectratio]{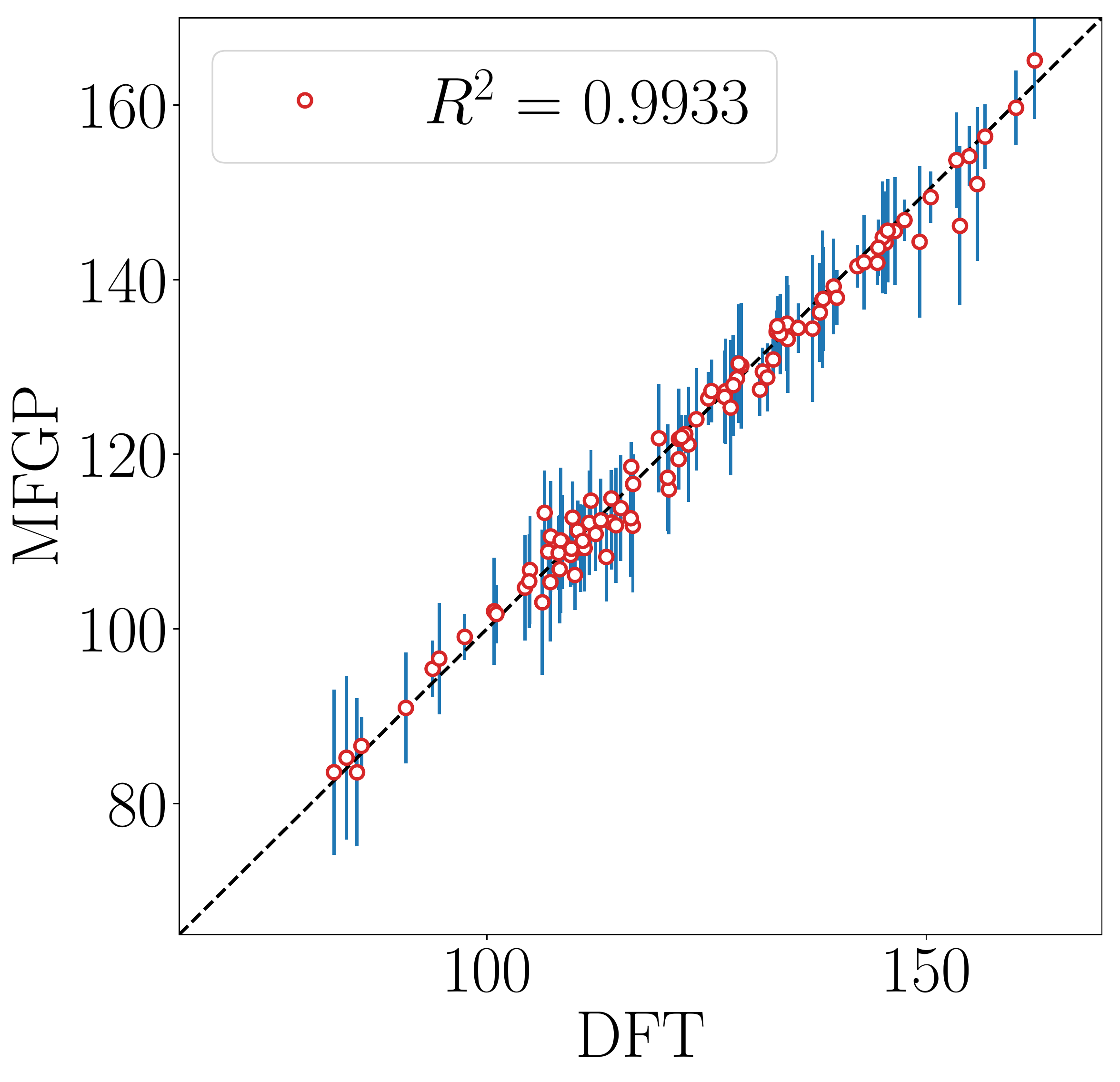}
	\caption{MFGP-50}
	\label{fig:cropped_predDFTvsMFGP-50-50}
\end{subfigure}
\hfill
\begin{subfigure}[b]{0.30\textwidth}
	\includegraphics[width=1\textwidth,keepaspectratio]{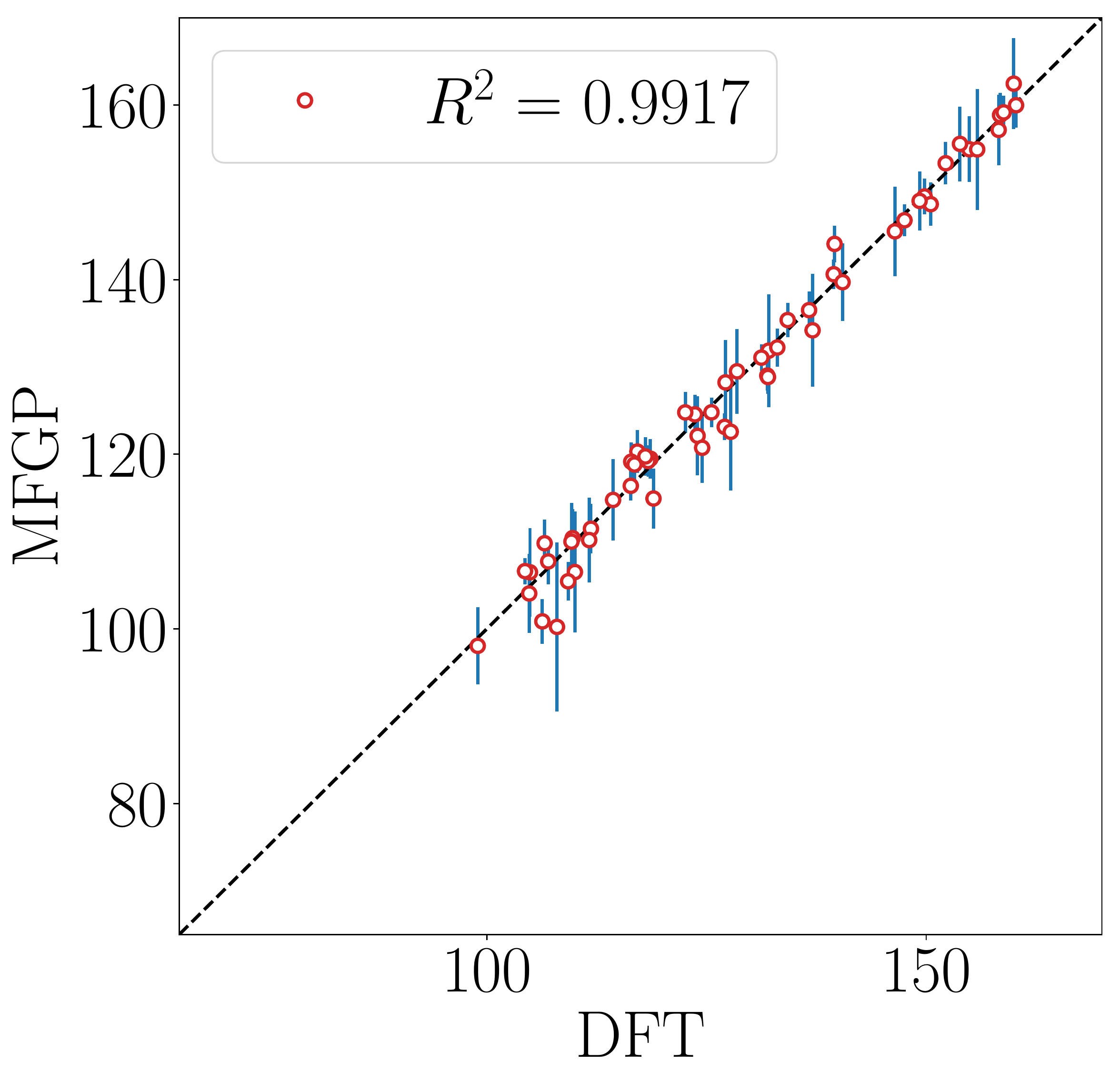}
	\caption{MFGP-70}
	\label{fig:cropped_predDFTvsMFGP-70-30}
\end{subfigure}
\hfill
\begin{subfigure}[b]{0.30\textwidth}
	\includegraphics[width=1\textwidth,keepaspectratio]{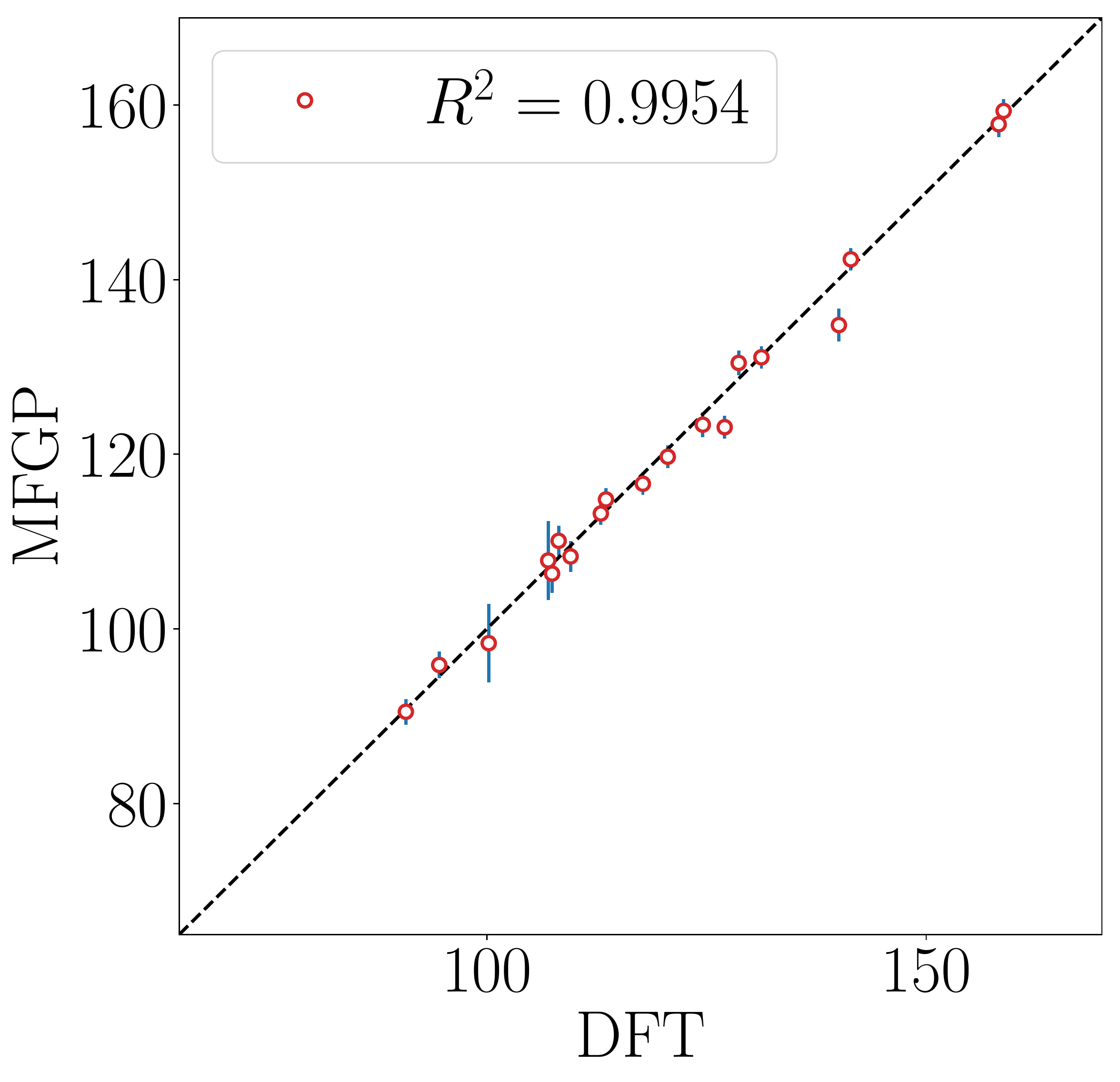}
	\caption{MFGP-90}
	\label{fig:cropped_predDFTvsMFGP-90-10}
\end{subfigure}
\hfill

  \caption{Comparison of the low-fidelity (SNAP, Figure
  \ref{fig:cropped_predDFTvsSNAP}) 
  and multi-fidelity 
  (MFGP, Figs.~\ref{fig:cropped_predDFTvsMFGP-10-90}-
  \ref{fig:cropped_predDFTvsMFGP-90-10}) models against the 
  high-fidelity 
  model (DFT) estimates of bulk modulus (GPa).
  For the SNAP model, estimates for all the AlNbTi ternary 
  compositions in Figure~\ref{fig:cropped_hfTernary0K} are shown.
  Pure aluminum, pure niobium, pure titanium and the equicomposition
  point for the grey, green, blue and black dots, respectively.
  For each MFGP models, the random fraction of the DFT data used
  in training is indicated in the subcaption \review{(MFGP-N 
  meaning that N \% of the HF dataset was used to train the 
  corresponding MFGP model)}, with remaining DFT points shown in the
  figure. Vertical blue bars indicate the associated uncertainty 
  quantification $\mu \pm 1.0 \sigma$.
  The legends indicate the squared correlation coefficients $R^2$.
}
\label{fig:ComparisonSnapMfgpDft}
\end{figure*}

Fig.~\ref{fig:ComparisonSnapMfgpDft} assesses the accuracy evolution of 
the approach by displaying the numerical predictions of the SNAP ML-IAP 
and of the MFGP results at various sizes of training 
datasets, with respect to the DFT predictions. For the
MFGP, it also
provides us with a measure of uncertainty quantification by
displaying the posterior Gaussian process prediction of 
$\mu \pm 1.0\sigma$ (vertical bars). 

Across the full ternary composition range, the SNAP potential predictions 
correlate with the DFT results with  $R^2 = 0.7122$ 
(Fig.~\ref{fig:cropped_predDFTvsSNAP}). 
Fig.~\ref{fig:cropped_predDFTvsSNAP} highlights the SNAP bulk modulus 
results of the pure element and equicomposition points. 
As explained in section~\ref{subsec:MDsimulations},  about 70\% of the 
training set of the SNAP potential consisted in data at the
equicomposition point, which explains the very good agreement observed 
for that particular composition. 
Figs.~\ref{fig:cropped_predDFTvsMFGP-10-90}-\ref{fig:cropped_predDFTvsMFGP-90-10} display the rapid
convergence of the MFGP towards the DFT results with an increasing
training dataset size.
The dataset  size was varied by randomly choosing a fraction of the full HF dataset. 
The remaining HF points were used as a testing dataset and are plotted in Figs~\ref{fig:cropped_predDFTvsMFGP-10-90}-\ref{fig:cropped_predDFTvsMFGP-90-10}.

With a 10\% training dataset (19 LF and 19 HF data points, Fig.~\ref{fig:cropped_predDFTvsMFGP-10-90}) to train the MFGP 
model, the MFGP performs on a par with the LF SNAP model ($R^2 =
0.7235$). 
Increasing the training dataset to 30\% (57 LF and 57 HF data points,
Fig.~\ref{fig:cropped_predDFTvsMFGP-30-70}) and 50\% (95 LF and 95 HF 
data points, Fig.~\ref{fig:cropped_predDFTvsMFGP-50-50})
considerably improved the accuracy of the prediction ($R^2 =
0.9813$ and $R^2 = 0.9933$, respectively) and reduced the
uncertainty. 
Additional increases to 70\% (133 LF and 133 HF data points, 
Fig.~\ref{fig:cropped_predDFTvsMFGP-70-30}) and 90\% (171 LF and 171 HF 
data points, Fig.~\ref{fig:cropped_predDFTvsMFGP-90-10}) of the
training dataset do not significantly improve the predictions
(i.e. the posterior mean) as there is almost no change in the 
correlation coefficients $R^2$, but increase the confidence in
those predictions by reducing the uncertainty (i.e. the posterior 
variance). 

\begin{figure}[!htbp]
    \centering
    \includegraphics[width=0.8\columnwidth,keepaspectratio]{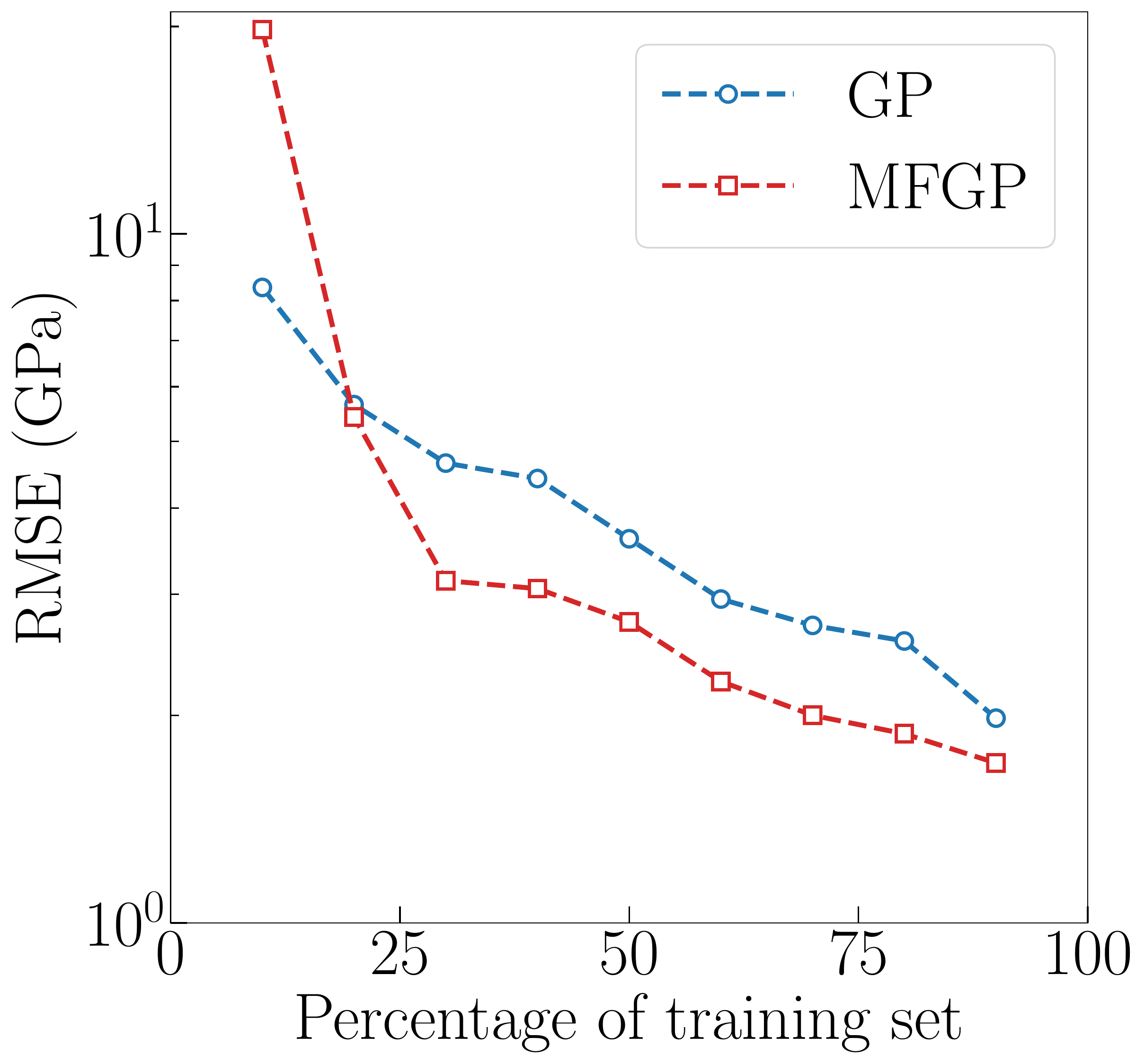}
    \caption{\review{Convergence of the GP (single-fidelity) and MFGP (multi-fidelity) RMSE of bulk modulus prediction as a function of the amount of training data used. For each point, the HF training and testing training data are the same. }}
    \label{fig:cropped_rmseConvergence}
\end{figure}

\review{In order to probe the improvement in learning efficiency 
obtained from the use of the MFGP model, 
we generated learning curves 
for both MFGP and single-fidelity GP. 
Fig.~\ref{fig:cropped_rmseConvergence} compares the bulk modulus 
root mean-squared error (RMSE) convergence of the MFGP to a 
single-fidelity GP model.
Once a sufficient amount of data is added to the training set, the RMSE 
of the MFGP model is consistently below that of the single-fidelity GP.
This demonstrates the improved learning obtained by adding a level 
of fidelity to the model.
}

\review{
A computational cost assessment and comparison of the 
two levels of fidelity and of the MFGP model was also performed. 
For consistency, all calculations were performed on Sandia's
\emph{Solo} cluster consisting of
dual-socket Intel Xeon E5-2695 (Broadwell) CPUs with 36 cores per 
node, and an Intel Omnipath interconnect.
Each HF and LF bulk modulus prediction is obtained in computation times of
approximately 4.1 and 1.83$\times 10^{-6}$ hours, on 2 and 1 \emph{Solo}
nodes, respectively.
This leads to relative speeds of 297.6 hours per core for the HF model, 
and of 6.6 $\times 10^{-5}$ hours per core for the LF model.
In comparison, the MFGP model performs a bulk modulus prediction with 
a relative speed of 1.95$\times 10^{-7}$ hours per core (approximately
1.95$\times 10^{-7}$ hours to perform one bulk modulus prediction on one 
core).
The relative speed of each different model (HF, LF and MFGP) differs 
by orders of magnitude. Once trained, the MFGP model allows to perform 
predictions with accuracy comparable to the HF calculations for an 
almost negligible computational cost. 
}

\subsection{Multi-fidelity Bayesian optimization of bulk modulus}

\begin{figure*}[!htbp]

\centering
\begin{subfigure}[b]{0.45\textwidth}
	\centering
	\includegraphics[width=1\textwidth,keepaspectratio]{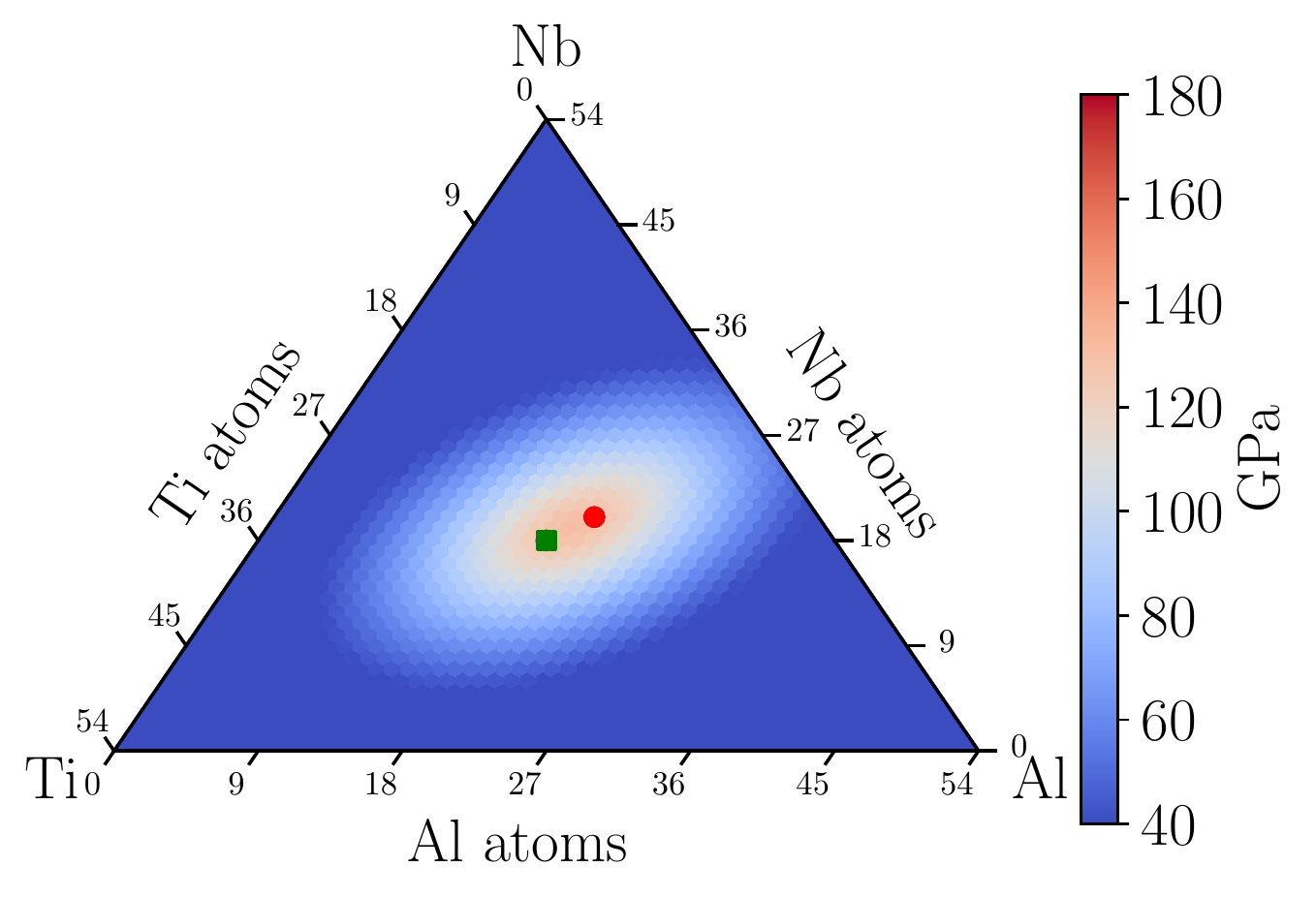}
	\caption{Iteration 4: 2 LF + 2 HF.}
	\label{fig:cropped_mfbo_iter5}
\end{subfigure}
\hfill
\begin{subfigure}[b]{0.45\textwidth}
	\centering
	\includegraphics[width=1\textwidth,keepaspectratio]{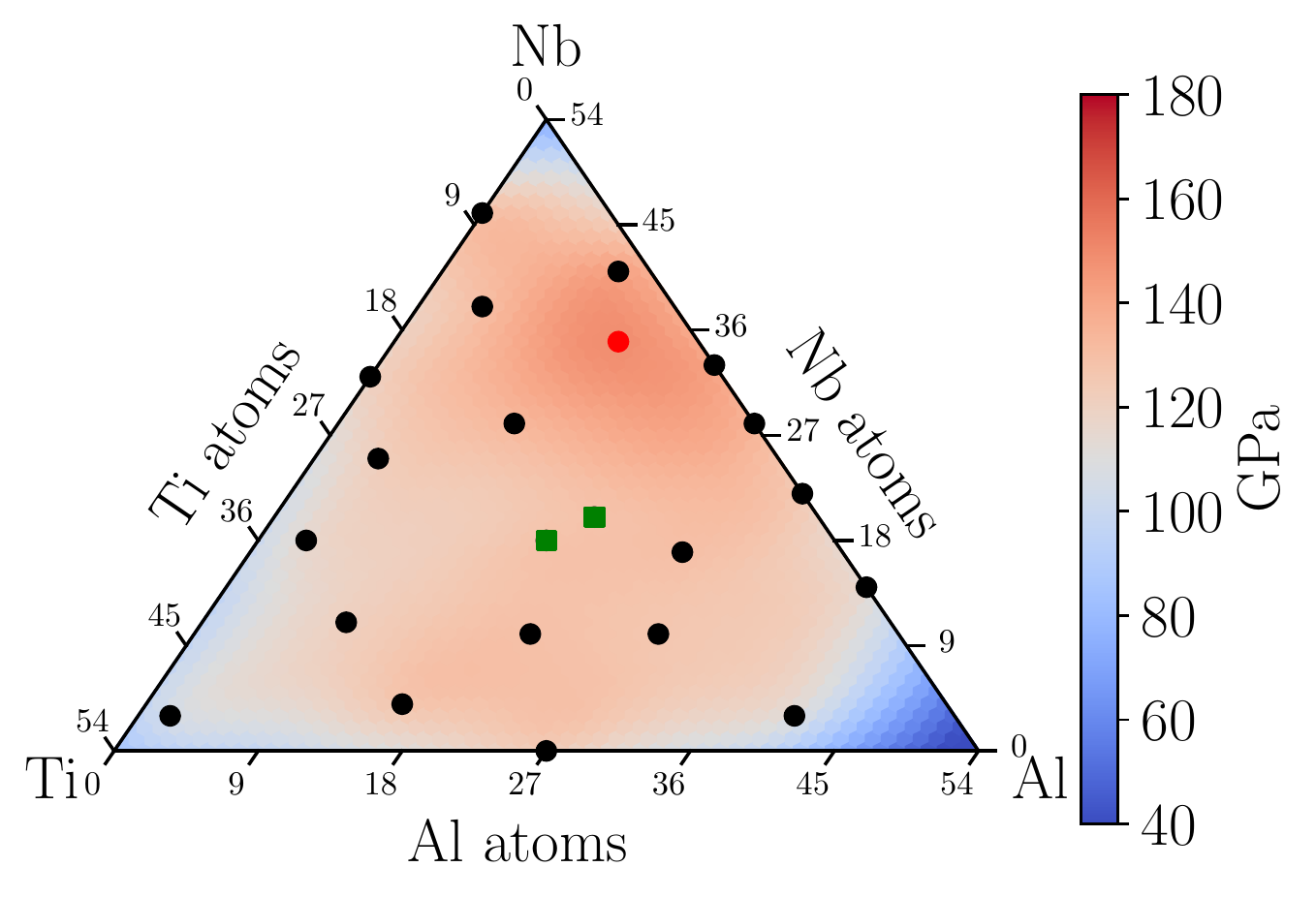}
	\caption{Iteration 24: 21 LF + 3 HF.}
	\label{fig:cropped_mfbo_iter25}
\end{subfigure}
\vfill

\begin{subfigure}[b]{0.45\textwidth}
	\includegraphics[width=1\textwidth,keepaspectratio]{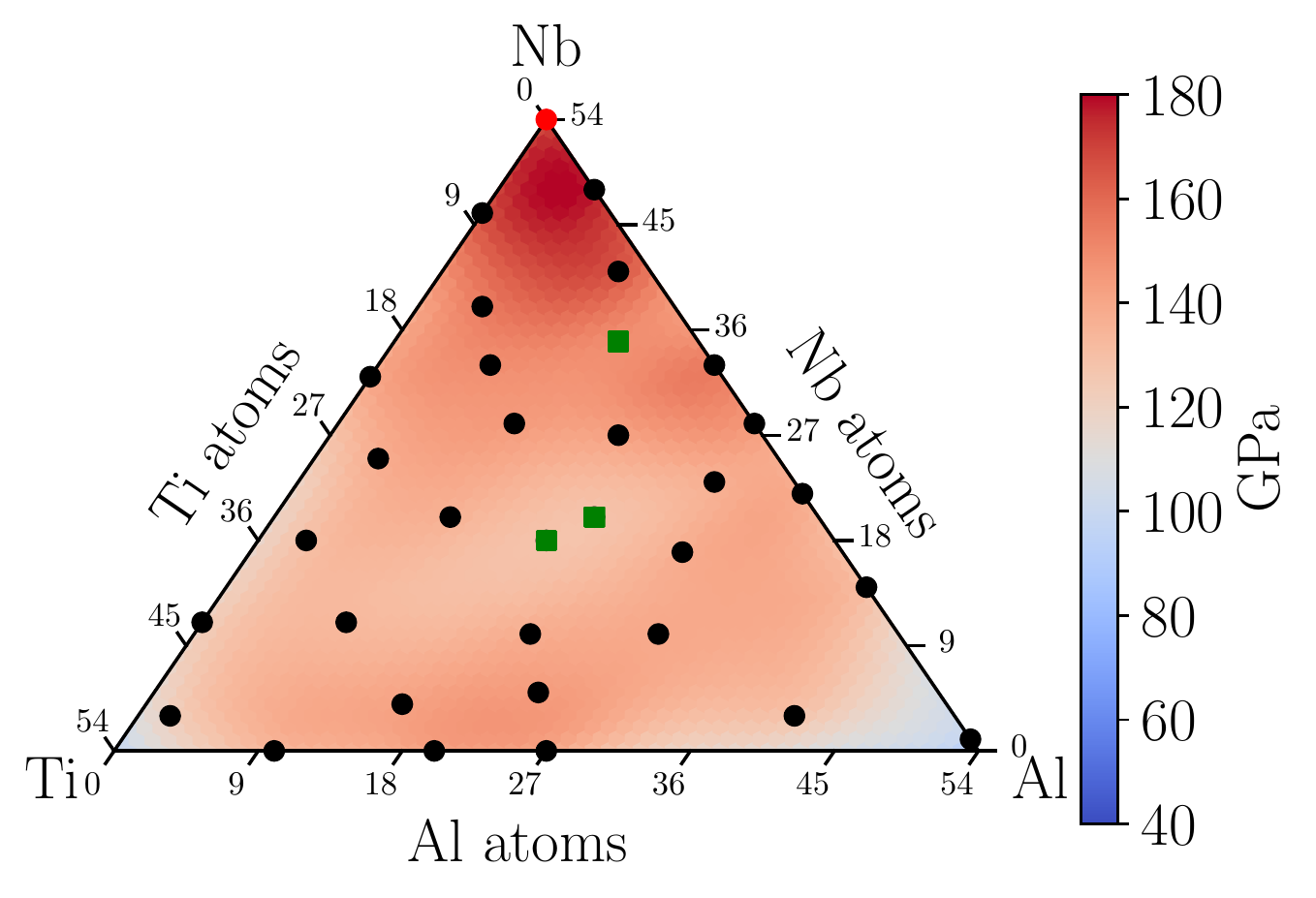}
	\caption{Iteration 35: 31 LF + 4 HF.}
	\label{fig:cropped_mfbo_iter36}
\end{subfigure}
\hfill
\begin{subfigure}[b]{0.45\textwidth}
	\includegraphics[width=1\textwidth,keepaspectratio]{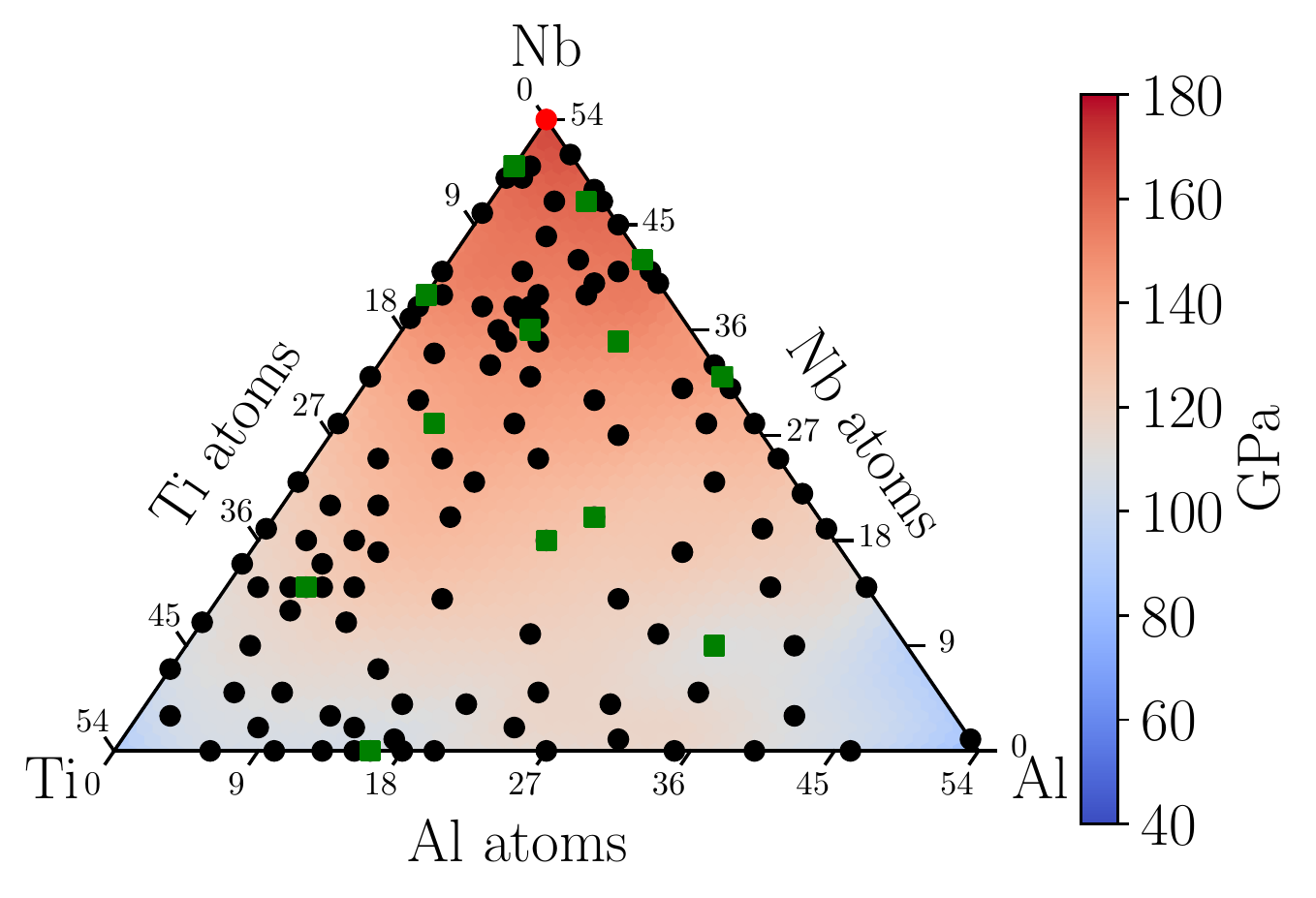}
	\caption{Iteration 130: 116 LF + 14 HF.}
	\label{fig:cropped_mfbo_iter131}
\end{subfigure}

\hfill

\caption{Progression of the multi-fidelity Bayesian optimization (MF BO) 
  for bulk modulus in the ternary composition space of AlNbTi alloys. The 
  color map indicates the MFGP predictions at iterations 4, 24, 35, and 130. 
  The red dot indicates the best HF data point evaluated so far, while the 
  green squares indicate the other HF data points and the green black dots 
  indicate LF data points.  At iteration 35, MF BO finds the global optimum 
  near the niobium vertex after evaluating only 4 HF points.
}
\label{fig:MFBO}
\end{figure*}

We demonstrate the application MF BO for practical
materials design by searching for the chemical composition
optimizing the QoI across the ternary AlNbTi composition range, and
using the MFGP framework to couple the SNAP and DFT predictions. 
The computational cost ratio between DFT and MD simulations is set 
to 10.0, and the UCB acquisition function is used.

The concept of using the acquisition function described in 
Section~\ref{subsec:AcqFunc} allows the MF BO to sequentially sample at the most informative point. 
Four sampling points, including 2 LF and 2 HF points around the 
equicomposition point, are used to build the initial MFGP, as shown 
in Fig.~\ref{fig:cropped_mfbo_iter5}. 
At iteration 24, the MF BO queries another HF evaluation 
(Fig.~\ref{fig:cropped_mfbo_iter25}) in the vicinity of the global 
maximum given by the LF predictions, as shown in 
Fig.~\ref{fig:cropped_lfTernary0K}, and corrects itself after 
obtaining the corresponding HF sampling value. 
The global maximum is obtained at iteration 35 
(Fig.~\ref{fig:cropped_mfbo_iter36}) after only four HF 
and 31 LF evaluations. 



\section{Conclusions}
\label{sec:Conclusion}

This work presented and applied a scale-bridging MFGP framework 
fusing information from DFT and MD calculations performed with a
SNAP ML-IAP. 
Equation-of-state (EOS) calculations were performed with DFT and 
the SNAP potential across the full AlNbTi ternary composition range. 
The EOS data allowed us to extract 190 high- and 1540 low-fidelity 
predictions of the QoI, the bulk modulus.
This full dataset (HF + LF predictions) is used as a 
training set. The HF configurations are also used as reference 
values to probe the validity of our approach (testing set).

The MFGP framework is then applied to this dataset to build a MF
model.
Its efficiency is demonstrated through the construction of a
high-resolution and highly accurate ternary composition diagram for the QoI
using only 50\% of the HF data points
(corresponding to 95 EOS evaluations performed with DFT) with an 
excellent correlation coefficient of $R^2 = 0.9933$ and low prediction uncertainty.
Additional increase of the training dataset size allowed even lower
prediction uncertainty.

By leveraging LF-HF correlations, our framework drastically
reduces the amount of expensive \emph{first principles}
calculations necessary to obtain a dense and  highly accurate ternary composition diagram for the property of interest. 
An MF BO algorithm is finally presented and tested by performing an
optimization of the QoI. 
After only 4 HF and 31 LF evaluations, the MF BO algorithm was able 
to locate the QoI optimum value across the composition space. 
Performed on the fly, the computational cost associated with materials 
property optimization and design would be drastically reduced.

For the QoI we used in this study, the optimization problem 
was trivial, as pure niobium has the highest value of bulk modulus
across the AlNbTi composition space.
However, reapplying the same framework to a different material 
composition space and QoI is straightforward. 
In future work, our framework will be extended to perform 
multi-objective \cite{tran2020srmobo3gp,shu2020new},  
constrained mixed-integer \cite{tran2019constrained} MF calculations,
allowing to  search for optimum compromises between different materials
properties in an asynchronously parallel manner
\cite{tran2019pbo,tran2020an}. 

Our results demonstrated the efficiency of our a MFGP framework to fuse in 
the predictions and therefore bridge the gap between two of the most commonly
employed atomistic simulation approaches (DFT and molecular dynamics), each 
of them acting at very different length and time scales. 
The same methodology remains valid for different scales, and could be 
leveraged to build materials modelling 
road-maps~\cite{van2020roadmap} by understanding the 
existing correlations between atomistic methods
\cite{bulatov1998connecting,tranchida2018massively,zepeda2017probing}
and associated higher scale coarse-grained~\cite{tranchida2018hierarchies} or continuum numerical models~\cite{arsenlis2007enabling,roters2010overview}.

\section*{Acknowledgment}

A.T. and J.T. contributed equally to this work. 
This research was supported by the U.S. Department of Energy, Office 
of Science, Early Career Research Program, under award 17020246 
and Sandia LDRD program 219144. 

The views expressed in the article do not necessarily represent the 
views of the U.S. Department of Energy or the United States Government. 
Sandia National Laboratories is a multimission laboratory managed and 
operated by National Technology and Engineering Solutions of Sandia, 
LLC., a wholly owned subsidiary of Honeywell International, Inc., 
for the U.S. Department of Energy's National Nuclear Security 
Administration under contract DE-NA-0003525. 

\section*{Data Availability}

The data that support the findings of this study are available
from the corresponding authors upon reasonable request.

\bibliographystyle{apsrev4-1}
\bibliography{lib}

\end{document}